\documentclass[preprint,showpacs,floats,letterpaper,floatfix,,groupedaddress,eqsecnum]{revtex4}
\usepackage{subfigure}

\usepackage{amssymb,amsmath}
\usepackage[dvips]{graphicx}

\usepackage{dcolumn,epsfig}

\begin{document}

\title{Dynamical States, Stochastic Resonance and Ratchet Effect in a 
Biharmonically Driven Sinusoidal Potential}
\author{W.L. Reenbohn and Mangal C. Mahato}
\email{mangal@nehu.ac.in}
\affiliation{Department of Physics, North-Eastern Hill University,
Shillong-793022, India}

\begin{abstract}
Two stable dynamical states of trajectories of an underdamped particle, under 
appropriate conditions, appear naturally in a sinusoidal potential when driven 
by a low amplitude biharmonic external field. These states are quite stable at 
low temperatures but make transitions between them as the temperature is 
raised. The proper choice of the biharmonic drive makes it possible for the 
system to show, at the same time, {\it{both}} the phenomena of stochastic 
resonance and ratchet effect. Ratchet effect, in this case, a consequence of 
the biharmonic drive, is obtained over a large domain of parameter space. 
However, stochastic resonance can be obtained only over a restricted 
(sub)domain of parameter space and owes its existence largely to the existence 
of the two dynamical states.
\end{abstract}

\vspace{0.5cm}
\date{\today}

\maketitle

\section{Introduction}
The phenomenon of stochastic resonance (SR) was discovered\cite{Benzi} almost 
a decade prior to some innovative experiments\cite{Svoboda} that brought the 
ratchet effect (RE) to renewed attention. SR is a phenomenon that shows a 
maximum in the response of a nonlinear system to a weak (subthreshold) input 
signal of a given frequency as the strength of noise in the input signal (or 
in the system) is varied. There has been intense investigations on the subject 
both theoretically and experimentally and have been reviewed 
extensively\cite{Gamma,Well}. However, the developments have been confined 
mostly to {\it{bistable systems}}; some of the early examples include the 
model Landau potential (theory), Schmitt trigger circuit, two-mode ring lasers
(experiment), etc\cite{McN, Fauve, Roy}. In sharp contrast, RE is a phenomenon 
in which a net current is obtained in {\it{periodic potential systems}} without 
application of any apparent bias, albeit, again in the presence of noise or
zero-mean fluctuating forces. The subject of study (RE) since had a phenomenal 
growth\cite{Julicher,Reimann,Mahato}. Both the phenomena were hailed as 
important scientific developments with harmonious consequences\cite{Maddox}. 

In the beginning of the current resurgence of investigations RE was put forth  
as an important model to explain material transport along microtubules in 
biological systems. There are essentially two prominent models of RE; other
models can be considered as variants of these two but are important on their 
own right. In one of the two models (rocked ratchet) the mean slope of the 
periodic potential is changed continuously (or abruptly) between positive and 
negative values\cite{Magnasco}. This is the model we are interested in this 
work. In the other important model (of flashing ratchets) an asymmetric 
periodic potential of fixed strength is switched on and off either randomly 
or periodically\cite{Prost}. Based on the latter model, RE was soon 
demonstrated experimentaly in a colloidal particle system\cite{Rousselet}. 
Both the phenomena (SR and RE), however, are counterintuitive and operate 
under nonequilibrium conditions in presence of noise.

In order for both the phenomena of SR and RE to occur in the same system, a 
driven periodic potential system needs to exhibit SR. As mentioned earlier the
study of SR is largely confined to bistable or two-well systems. The occurrence 
of SR in periodic potential systems has not been conclusively proved 
analytically. 
However, some recent numerical investigations unmistakably indicate existence 
of SR in underdamped periodic potential systems too\cite{Saikia,Wanda}. This 
opens up the possibility of seeking both the phenomena to occur in the same 
system under the same conditions. The present work shows that the simultaneous 
occurrence of SR and (rocked) RE is a distinct possibility even in case of 
underdamped symmetric periodic potential systems. It has been found earlier 
that SR and RE occur simultaneously in an overdamped flashing ratchet where 
the periodic potential is considered asymmetric\cite{Qian}. As far as the 
present authors are aware that is the only report where RE and SR are 
genuinely found to occur simultaneously and the two phenomena are shown to have
a close relation. 

In that work\cite{Qian}, the power spectrum of the cosine of the position 
$x(t)$ for all $t\ge 0$ gives a dominant peak at the potential switching 
frequency at non-zero temperatures. This peak shows a maximum at a temperature 
very close to where the ratchet current has the largest value. Since the 
maximum of the power spectrum peak shows SR it was concluded that there is a 
close connection between SR and RE. Curiously, at zero temperature the power 
spectrum shows strong peaks at frequencies about one third (and multiples 
thereof) of the potential switching frequency. These intra-(potential)well 
breathings seem unrelated to switching field and disappear as the temperature 
is raised. In contrast, in the present work the frequency of external field is 
necessarily taken to be close to the natural frequency of oscillation at the 
bottom of the potential wells. Also, we do not really find any strong evidence 
of a close relationship between the two phenomena.
   
Periodic potentials are almost ubiquitous and stochastic motion of particles
along periodic potentials have been studied extensively\cite{Risken}. The 
examples include, crystals, semiconductor heterostructures, RCSJ model of 
Josephson junctions, microtubules, etc. Therefore, the occurrence of optimised 
signal (SR) and material transport (RE) together seems a very attractive 
possible proposition and needs to be investigated.

It has been shown earlier\cite{Saikia,Wanda} that when an underdamped particle, 
subjected to temporally periodic (sinusoidal) field, moves in a spatially 
periodic (sinusoidal) potential, the trajectory $x(t)$ of the particle shows 
periodic behavior at low temperatures. The amplitude of $x(t)$ and its phase 
relationship with the exernal field $F(t)$ depend on the amplitude and 
frequency of $F(t)$, the friction coefficient $\gamma$ and also on the initial 
conditions of position $x(0)$ and velocity $v(0)$. A small range of frequencies 
($\omega=\frac{2\pi}{\tau}$) of $F(t)$ close to the natural frequency 
($\omega_0=1$) of free oscillation at the bottom of the sinusoidal potential 
have a remarkable influence on the nature of trajectories.

In a narrow range around $\tau=2\pi$ of the period $\tau$ of $F(t)$ two 
distinct kinds of trajectories $x(t)$ are realized: one having a small phase 
lag $\phi$ with respect to $F(t)$ and the other with a rather large $\phi$ and 
amplitude. These two trajectories $x(t)$ genuinely have the status of dynamical
states having distinct basins of attraction in the ($x(0)-v(0)$) phase space. A
particle in the in-phase (small $\phi$) state usually dissipates on the average
much less energy per period, 
\[\overline{W}=\int{x(t)dF(t)}\]
than in the out-of-phase (large $\phi$) state. For a given value of $\tau$ and
the friction coefficient $\gamma$ the relative abundance of the states in the 
(x(0),v(0)=0) space over a period of the potential $V(x)$ (for example, 
[$-\frac{\pi}{2}\le x(0)<\frac{3\pi}{2}$] for $V(x)=-\sin(x)$) and also their 
relative stability depend on the amplitude $F_0$ of $F(t)$. For a given set
of $\tau$, $\gamma$, and $F_0$, as the temperature is gradually increased,
the relative abundance of the two dynamical states also change on the average
and conspire together in such a way that the overall average energy dissipation
$<\overline{W}>$ shows a maximum. In the periodic potential systems 
$<\overline{W}>$ is preferable and appropriate as a quantifier of SR over the
signal to noise ratio\cite{SRS}. Thus it is shown that SR is obtained in a 
periodic potential system but is explained in terms of transitions between the 
two dynamical states just as in the case of bistable systems. It may, however,
be noted that in the conventional SR in double-well potential systems 
transition between the two states takes place at frequencies close to the 
Kramers rate of passages between the two wells. The frequencies of the drive 
field $F(t)$ in the bistable systems, therefore, turn out to be about a factor
of $10^{-2}$ of the frequencies used in the present study. In the present work 
we consider a biharmonic external field $F(t)$ with the main component of 
frequency $\omega$ close to $\omega_0$ and an additional component either of 
frequency $2\omega$ (harmonic) or $\omega/2$ (subharmonic).

\begin{figure}[htp]
\centering
\includegraphics[width=15cm,height=10cm,angle=0]{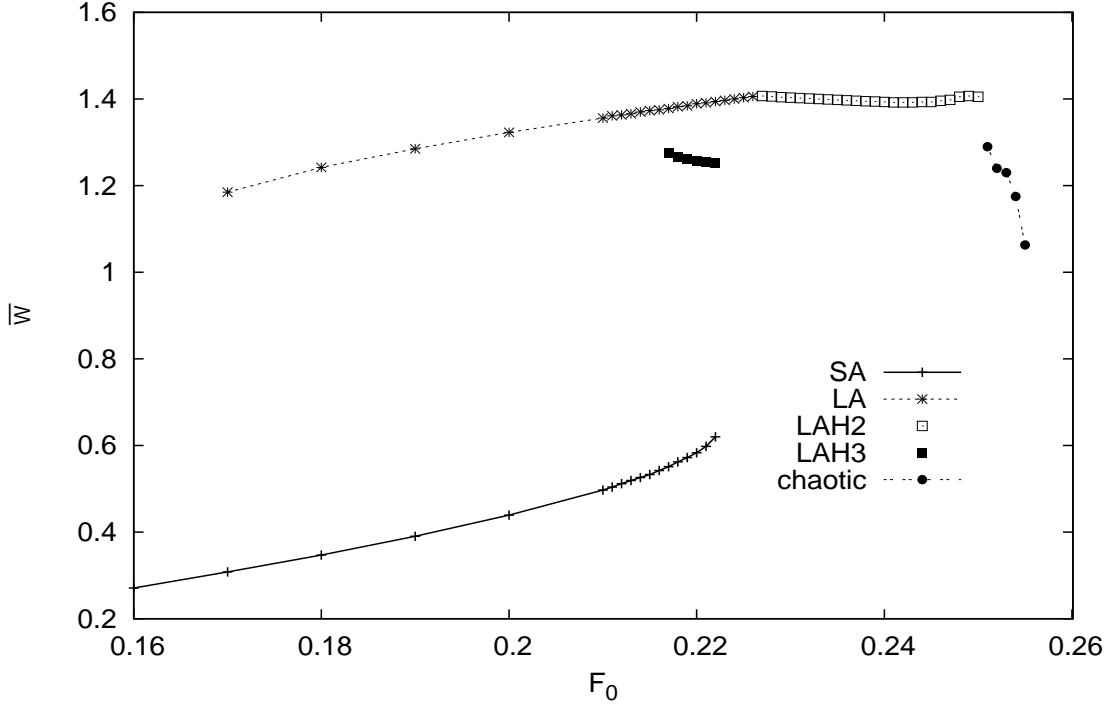}
\caption{The figure shows the input energies $\overline{W}$ of various
dynamical states SA, LA, LAH2, LAH3 and chaotic, occuring when
 driven by a biharmonic drive of amplitude $F_0$, $f_m=0.2$,
 the period of the main frequency component $\tau=8.0$ at
 $T=0.000001$.}
\end{figure}

\begin{figure}[htp]
\centering
\includegraphics[width=10cm,height=15cm,angle=0]{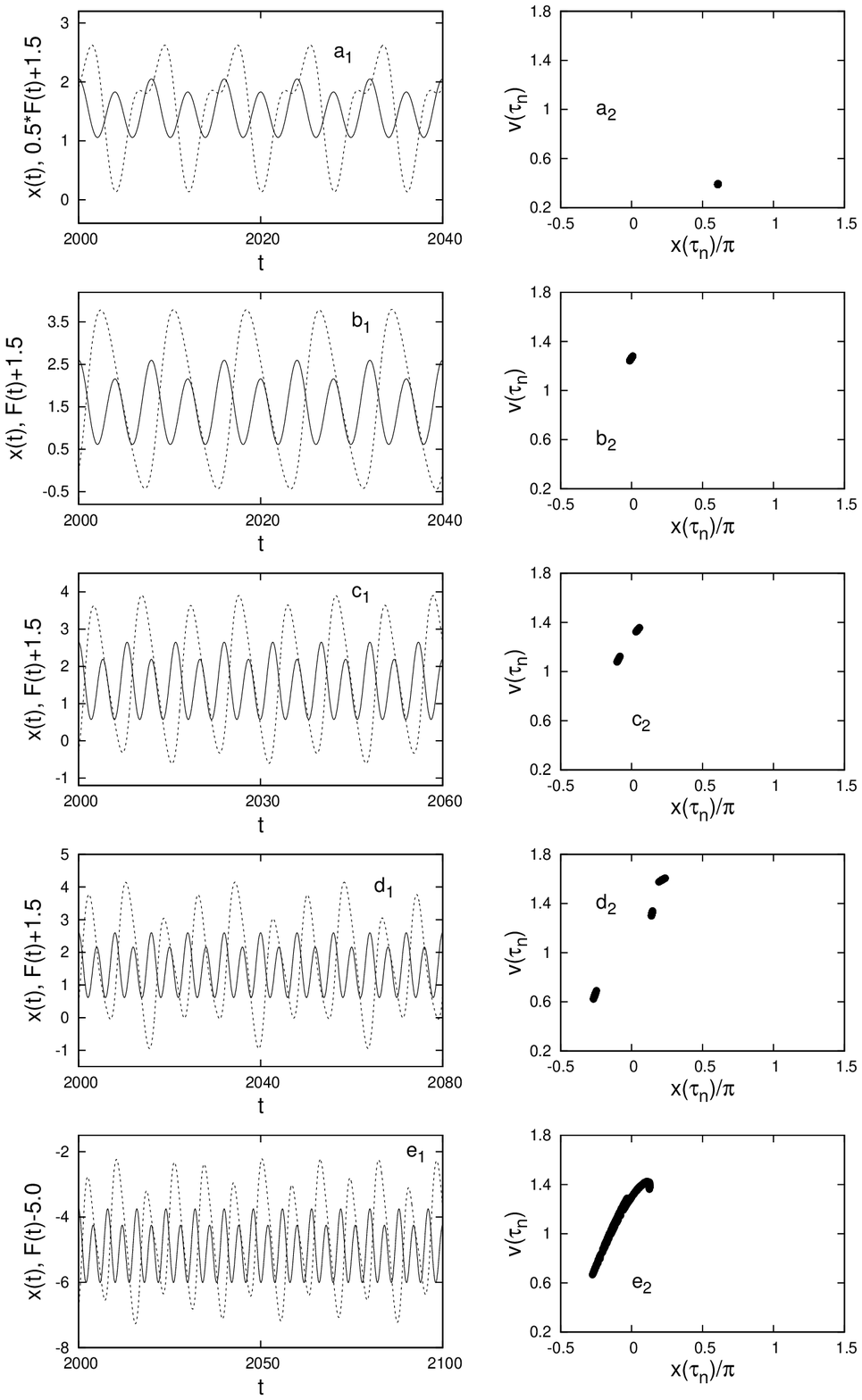}
\caption{The particle trajectories (dotted lines) of states SA, LA, LAH2, LAH3 
and chaotic states of Fig.1. For comparison the field $F(t)$ are also plotted
(continuous line) along with the trajectories. The corresponding stroboscopic  
plots are also shown on the adjacent panel as Figs. 2 a$_2$, b$_2$, c$_2$, 
d$_2$ and e$_2$ at $t=\tau_n=n\tau,~n=0,~1,~2,\cdots$.}
\end{figure}

As in the case of pure sinusoidal drive (described above), when the underdamped 
particle in the sinusoidal potential is driven by a biharmonic (harmonic or 
subharmonic) field similar dynamical states are realized. However, now the 
states become more complex and richer. As an illustration, the $\overline{W}$ 
calculated over a period $\tau$, for $\gamma=.12,~\tau=8.0$, the fraction of 
the main frequency term ($f_m=.2$) ($f_m$ is defined in the next section), 
and temperature $T=.000001$, are plotted as a function of $F_0$, the 
amplitude of the main frequency component of the drive in Fig.1. (All the 
parameters are given in dimensionless units, as described in the next section.) 
For convenience (and also to avoid the difficulty of assigning unambiguously 
the value of phase lag $\phi$), we name the dynamical states (based on the 
amplitude of trajectories) as small-amplitude (SA) or large-amplitude (LA) 
states instead of in-phase and out-of-phase states. 

As is indicated in Fig.1, we obtain three kinds of LA states in the harmonic
drive case: LA (with period $\tau$), LAH2 (with period $2\tau$) and LAH3 (with 
period $3\tau$). These LA states are obtained in different regimes of $F_0$ 
and have different $\overline{W}$. If the value of $F_0$ is increased beyond 
0.25, the trajectories no longer remain periodic but become chaotic even at as 
low a temperature as .000001. The representative trajectories, along with 
their stroboscopic plots, corresponding to these states are shown in Fig.2. 

The relative stability and hence the relative abundance of these states 
existing at a given temperature decide the value of $<\overline{W}>$ at that 
temperature\cite{Wanda}. The maximum shown by $<\overline{W}>$ as a function 
of temperature indicates the occurrence of SR. However, the significance of 
such a peak, as a signature of SR, loses its meaning if it happens to occur at 
a temperature comparable to or higher than the potential barrier height 
itself. On the other hand, there could be situations, for example at large 
amplitudes of $F_0$, where $<\overline{W}>$ may peak sharply at such low 
temperatures that only intra-(potential)well transitions between states take 
place. These peaks cannot be considered as an indication of occurrence of SR
in a periodic potential. These restrictions make the domain of occurrence of 
SR in the parameter space quite narrow. However, RE has a comparatively large 
domain of occurrence.

From Fig.1, we find that there is a range of $F_0$ where the two states LA and 
SA coexist in the harmonic drive case. This is the range of $F_0$ where SR is 
found to occur. As the amplitude is increased, the states LAH3 also appear and 
one can also see peaking of $<\overline{W}>$ with temperature. However, this 
peaking of $<\overline{W}>$ cannot be genuinely ascribed to SR because it 
takes place at a low temperature with only intra-well transitions between the 
states. Thus, we find that SR is due to transitions between the states SA and 
LA even in this case of biharmonic drive and the dynamical states LAH3 and 
LAH2 do not contribute to SR.

In the next section, the mathematical model will be discussed. The biharmonic
drive parameters will be clarified. In Sec.3 the numerical method adopted and
results obtained will be described in detail. Finally, in Sec.4 the results 
obtained will be discussed and conclusions of the work will be given.

\section{The Model}
As usual, the motion of an underdamped particle of mass $m$ moving along
the one dimensional spacing $x$ in a sinusoidal potential $V(x)$ and driven by 
a time periodic forcing $F(t)$ subjected to a Gaussian thermal fluctuating 
force $\xi(t)$ of mean zero is given by the Langevin equation\cite{Risken}
\begin{equation}
m\frac{d^{2}x}{dt^{2}}=-\gamma\frac{dx}{dt}-\frac{\partial{V(x)}}{\partial
x}+F(t)+\sqrt{\gamma T}\xi(t).
\end{equation}
Here, $\gamma$ is the coefficient of frictional force experienced by the 
particle while in motion with velocity $\frac{dx}{dt}$ and is related to the 
fluctuating force through the fluctuation-dissipation theorem
\begin{equation}
<\xi(t)\xi(t^{'})>=2\delta(t-t^{'}).
\end{equation}

Here $T$ is the temperature (in energy units, $k_B=1$) of the heat bath with 
which the particle is in thermal contact. The potential is taken as
\begin{equation}
V(x)=-V_0 \sin(kx),
\end{equation}
and the external periodic force is considered biharmonic in nature
\begin{equation}
F(t)=F'_0(f_m\cos(\omega t)+f_h\cos(2\omega t+\theta))
\end{equation}
in case of harmonic field, and
\begin{equation}
F(t)=F'_0(f_m\cos(\omega t)+f_h\cos(\frac{\omega}{2} t+\theta))
\end{equation}
in case of subharmonic field. Here, $f_m$ is the fraction of the main
frequency ($\omega$) term and $f_h~(=1-f_m)$ is the fraction of the harmonic 
(subharmonic) frequency term. $F(t)$ can be rewritten, for example, as
\begin{equation}
F(t)=F_0(\cos(\omega t)+\alpha\cos(2\omega t+\theta)),
\end{equation}
where 
\begin{equation}
F_0=F'_0f_m,
\end{equation}
and the ratio
\begin{equation}
\alpha=\frac{f_h}{f_m}.
\end{equation}
For simplicity and convenience the equations are rewritten in dimensionless
units by setting $m=1$, $V_0=1$, $k=1$, with reduced variables denoted again 
now by the same symbols. Thus, the Langevin equation takes the form
\begin{equation}
\frac{d^{2}x}{dt^{2}}=-\gamma\frac{dx}{dt}
-\frac{\partial V(x)}{\partial x} +F(t)+\sqrt{\gamma T}\xi(t),
\end{equation}
where the reduced potential $V(x)=-\sin(x)$. We define the effective potential
\begin{equation}
U(x,t)=V(x) -xF(t),
\end{equation}
with $F(t)$ given, for example, by the equation of the same form
\begin{equation}
F(t)=F_0(\cos(\omega t)+\alpha\cos(2\omega t+\theta)),
\end{equation}
for the harmonic drive, and
\begin{equation}
F(t)=F_0(\cos(\omega t)+\alpha\cos(0.5\omega t+\theta)),
\end{equation}
for the subharmonic drive, but with the new dimensionless $F_0$ equal in 
magnitude to the fraction $\frac{F_0}{V_0k}$ of the old dimensioned quantities 
and similarly for $\omega$ and $t$, leaving, of course, the magnitude of 
$\omega t$ same, etc\cite{Desloge}. Notice that, in this dimensionless form, 
the natural frequency $\omega_0$ of oscillation at the bottom of any well of 
the potential equals 1, as mentioned in the Introduction. Therefore, so long 
as $\gamma<<1$ we are working in the underdamped case. We keep the period 
$\tau$ of $F(t)$ of the main frequency term such that 
$\omega~(=\frac{2\pi}{\tau})\approx \omega_0$ appropriate for the study of SR 
in the sinusoidal potential. 

Note that whatever be the values of the parameters $F_0,~\alpha,~\theta$, the
total impulse applied on the particle over a period 
\begin{equation}
I=\int F(t)dt
\end{equation}
is zero. For $\theta=0$ and $\pi$, for example, large forces are applied 
over a smaller period in one direction but small forces act over a longer 
period in the other direction. This kind of force profiles are known to result 
in ratchet effect\cite{Reimann}. Certain organisms, analogously, use power 
(swift) and reverse (slower) strokes of their flagella to aid locomotion. 
$\theta=\pi$ gives the same $<\overline{W}>$ as $\theta=0$, that is the same 
SR profile. Both $\theta$ values will show RE but they will result in opposite 
net current directions. In contrast to $\theta=0$ and $\pi$, for 
$\theta=\frac{\pi}{2}$, $F(t)$ has a profile like an asymmetric saw-tooth 
repeated periodically. Here same kind of forces act in both directions but the 
increasing and decreasing slopes of $F(t)$ are different. Such $F(t)$ profiles 
also are known to exhibit ratchet effect\cite{Reimann,Mahato}. In all cases 
friction plays an important role and so does the noise.

The detailed numerical solution of the Langevin equation and analysis thereof, 
as described in the following, brings out the essential features of SR and RE 
in the system resulting from the biharmonic drive.

\section{Numerical Results}

We solve the Langevin equation, Eqn.(2.2), numerically to obtain the position 
$x(t)$ of the particle as time $t$ progresses. The second-order Heun's method 
is suitably adapted\cite{Mannella} to solve the Langevin equation (2.2) as an 
initial value problem for the purpose. We choose integration-time-step size 
$\Delta t = 0.001$. We take the initial velocity $v(0)\equiv v(t=0)=0$ for all 
cases. The initial position $x(0)\equiv x(t=0)$ are chosen at 100 equispaced 
points $x_i,~i=1,2,\cdots,100$ between the two consecutive peaks, e.g., 
[$0\le x_i<2\pi$]. The input energy, or work done by the field on the system 
(or equivalently energy dissipated by the system to the environment), $W$, in 
a period $\tau$ of the external field $F(t)$, is calculated as\cite{Sekimoto}:
\begin{equation}
W(t_0,t_0+\tau)=\int_{t_0}^{t_0+\tau}\frac{\partial U(x(t),t)}{\partial t}dt,
\end{equation}
\[=-\int_{t_0}^{t_0+\tau}x\frac{\partial F}{\partial t}dt\]
\[=-\int_{F(t_0)}^{F(t_0+\tau)}xdF.\]
Therefore,
\begin{equation}
W(t_0,t_0+\tau)=-\oint xdF = A,
\end{equation}
where $A$ is the magnitude of the area of the hysteresis loop $x(F)$. The 
average input energy per period, $\overline{W}$, averaged over an entire 
trajectory spanning $N_1$ periods of $F(t)$, is
\begin{equation}
\overline{W}= \frac{1}{N_1}\sum_{n=0}^{n=N_1}W(n\tau,(n+1)\tau)=\overline{A},
\end{equation}
where $\overline{A}$ is the area of the mean hysteresis loop $\overline{x}(F)$.
In our calculation, typically, we solve Eqn. (2.2) for $N_1 = 10^5$ periods of 
$F(t)$ for each initial condition $x(0)$ and take an ensemble average over
all initial conditions to obtain $\overline{x}(t)$ and $<\overline{W}>$. The
ratchet current or the mean velocity $\overline{v}$ is calculated by dividing 
the mean distance travelled by the total run time. 

Before we proceed to investigate the effect of biharmonic field on the
occurrence of SR and RE it will be quite educative to see the effect of the
frequency $\omega~(=2\pi/\tau)$ of the pure sinusoidal drive. As has been 
indicated earlier, a pure sinusoidal drive cannot yield RE in a symmetric 
sinusoidal potential system. We, therefore, study the occurrence of the two 
dynamical states LA and SA which are responsible for SR and find the domain in 
the parameter space of ($\gamma-\tau$) where LA and SA occur simultaneously.

\subsection{Pure sinusoidal drive}

\begin{figure}[htp]
\centering
\includegraphics[width=15cm,height=10cm,angle=0]{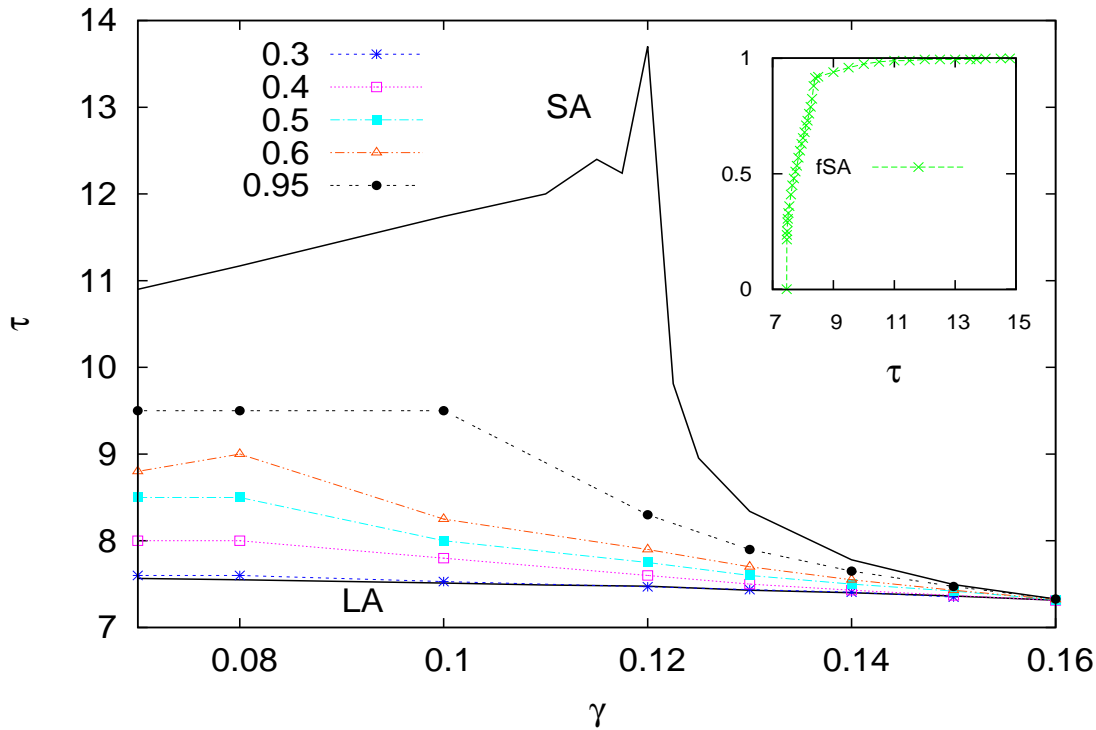}
\caption{The locii in the $\tau-\gamma$ plane of the fractions (fSA= 0, 0.3,
0.4, 0.5, 0.6, 0.95, and 1.0) of SA states in the coexistence region of the 
states SA and LA when the system is driven by a sinusoidal field of period 
$\tau$ and amplitude $F_0 =0.2$ in a medium of uniform friction $\gamma$. 
The inset shows the variation of fSA as $\tau$ is varied for $\gamma=0.12$. It 
shows a large discontinuous jump of 0.215 (from $0$) at $\tau=7.466$
and a very gradual and slow approach to fSA=1 at large $\tau$ values.}
\end{figure}

We restrict our study to the amplitude $F_0=0.2$ of the external field 
$F(t)=F_0\cos(\omega t)$ at $T=.000001$ for the occurrence of the 
states LA and SA. Note that, in the pure sinusoidal drive case, we set $f_m=1$. 
As mentioned earlier, we determine the nature of states (LA or SA) by looking
at the particle trajectories for 100 uniformly distributed initial positions
$x(0)$ drawn from inside of one period of the sinusoidal potential. 

Fig. 3 shows the diagram of existence of LA and SA states in the 
($\gamma-\tau$) plane. For a given $\gamma$ we find only LA state for 
small $\tau$ values. As $\tau$ is increased at a certain value of $\tau$ SA 
too begins appearing. This $\tau$ sets the boundary between regions of 
existence of pure LA states and the simultaneous occurrence of both SA and LA 
states. On further increase of $\tau$ the fraction of LA states decreases (and 
concurrently the fraction of SA states increases) and at a certain value of 
$\tau$ the fraction of LA states becomes zero for the first time. This value 
of $\tau$ sets the boundary between region of existence of the pure SA states 
and the region of coexisting states of LA and SA. Together with these extreme 
boundary lines of fractions 0 and 1 of the SA states are also plotted the 
locii of fractions .3, .4, .5, .6, and .95 of SA. As can be seen from the 
figure there is a large space between the locus of fraction .95 of SA and the 
boundary where LA disappears. This is because for certain small number of 
initial positions $x(0)$ the LA states continue to exist even for large values 
of $\tau$. This is particularly noticeable around $\gamma=.12$. However, as
explained below, the states appearing in this subspace do not make any 
significant impact on our study of SR.
 
In the inset of Fig. 3 are plotted the fractions (fSA) of states SA as $\tau$ 
is varied. The inset shows that the appearance of SA is very rapid and 100 
percent LA boundary stays close to $\tau=7.5$. However, the disappearance of 
LA is very slow and gradual and LA (together with SA) continues till a value 
of $\tau$ as large as 14. It also shows that there is a small range of $\tau$, 
close to $\tau=7.5$, where both the states exist in significant fractions. 
As will be clear in the following this remark has important bearing on SR for 
the biharmonic drive. Now, as the temperature (noise strength) is increased
from this small value transitions take place between the states SA and LA.
Consequently, the average input energy $<\overline W>$ changes as the
temperature is varied.

\begin{figure}[htp]
\centering
\includegraphics[width=12cm,height=12cm,angle=0]{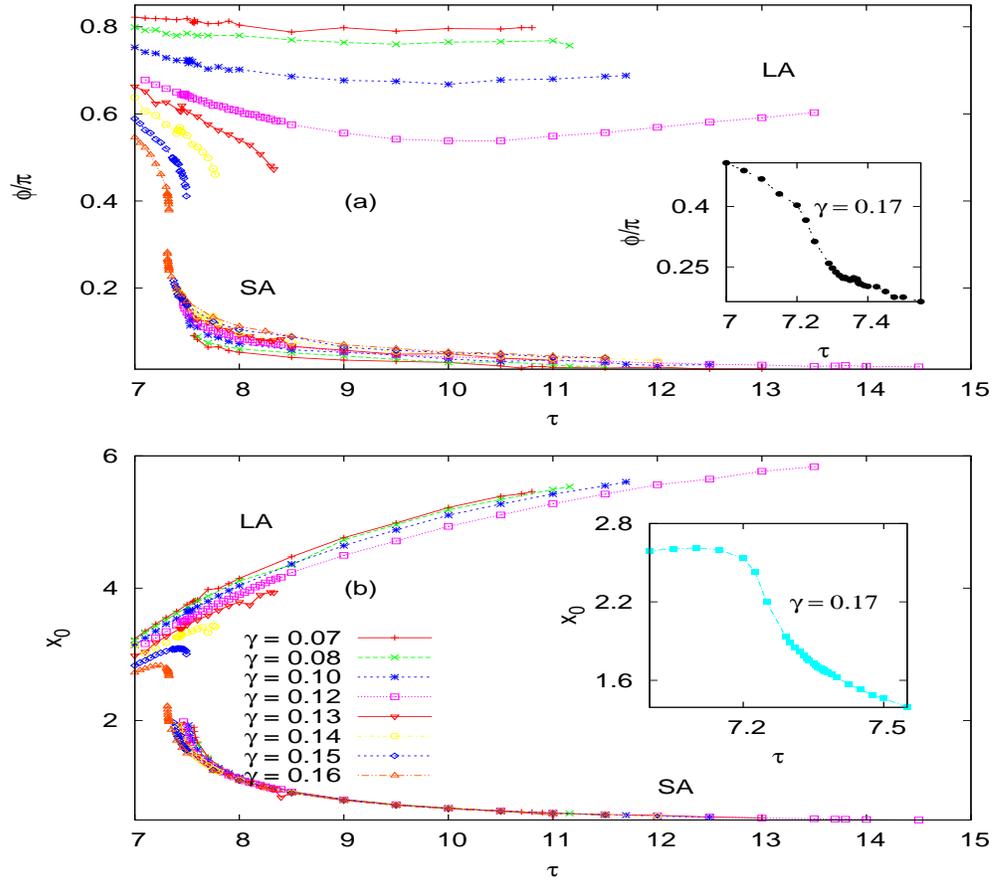}
\caption{Plot of the phase lags $\phi$ (a) and amplitudes $x_{0}$ (b) 
of the trajectories as a function of period $\tau$ of the drive for
various values of $\gamma$. We obtain two branches, for each $\gamma$, 
one correspond to SA and the other to LA. For each $\gamma$ there is
a region (the coexistence region) of $\tau$ where both SA and LA appear
but in separate branches. The inset shows that there is only one single 
continuous branch for $\gamma=0.17$.}
\end{figure}

Before we proceed to study SR it will be quite appropriate to appreciate the 
nature of these states SA and LA. These states make their appearance at $T=0$
and persist to temperatures $T$ much smaller than the potential barrier height 
of maximum 2 and driven by a subthreshold field of amplitude $F_0=.2$. In 
Fig. 4a we plot the phase lags $\phi$ between these states of mean trajectories
$\overline{x}(t)=x_0\cos(\omega t-\phi)$ and the drive field $F(t)$ and in 
Fig. 4b, their amplitudes $x_0$ are plotted as $\tau=\frac{2\pi}{\omega}$ is 
varied. We notice that both $\phi$ and $x_0$ change continuously as $\tau$ is 
varied for both the states for $\gamma$ values ranging from .07 to 0.16. The 
curves for SA and LA are clearly separated. The nature of variation of the 
curve for LA (SA) is just a continuation even at the boundaries where SA (LA) 
makes its appearance for the first time. A state, thus, either exists or does 
not exist at all and not that a state emerges out of the other state as $\tau$ 
is varied. It is clearly not a case of bifurcation at the boundary; these 
states have their independent identities. Bifurcations and associated chaos 
occur only at large $F_0$ values\cite{SaikiaPhysica}. For $\gamma\ge .17$, 
however, we get just one kind of trajectory whose $\phi$ and $x_0$ change 
continuously as $\tau$ is varied in the entire range (insets of Figs. 4). One 
cannot identify from the graph to call the trajectories SA or LA for 
$\gamma\ge .17$. 

\begin{figure}[htp]
\centering
\includegraphics[width=12cm,height=12cm,angle=0]{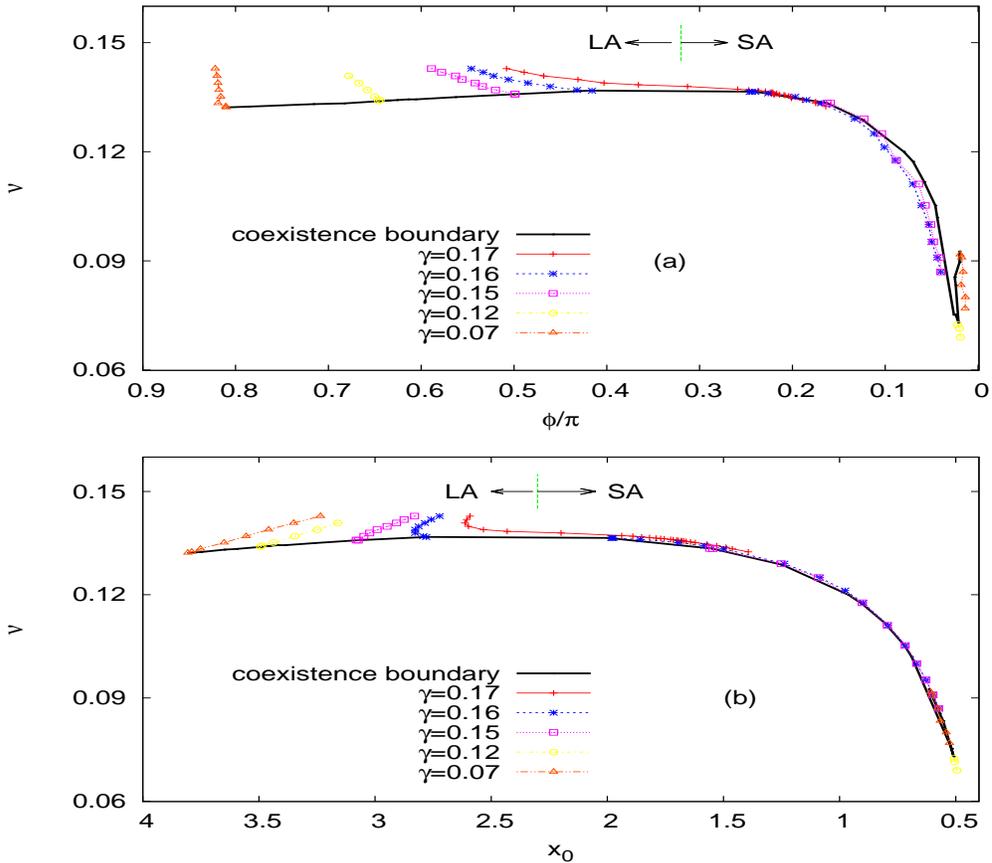}
\caption{Replot of Fig.4 of $\nu(=\frac{1}{\tau})$ versus $\phi$ (a)
and $\nu$ versus $x_{0}$ (b). The thick line shows the boundary
of the coexistence region of SA and LA. We have the SA and LA branches
outside the coexistence region of various $\gamma$ values.}
\end{figure}

In order to gain a little more insight into the dynamical states and their 
coexistence, the Figs. 4a and 4b are redrawn in Figs. 5a and 5b. In Fig. 5a the
frequency ($\nu=\frac{1}{\tau}$) of $F(t)$ is plotted against $\phi$ and in 5b
the frequency is plotted against the amplitude $x_0$ of the trajectories 
$x(t)$ for various constant $\gamma$ values. The boundary of the region of
coexistence of the two states is shown prominently leaving the region itself 
blank. The blank bounded region separates the two regions of occurrence of 
pure LA and SA states. An analogy may
be drawn with the liquid-gas pressure-density phase diagram with the
isothermals replaced by the constant $\gamma$ curves of Figs. 5. The SA states
('phases'), of course, run close to the boundary. In the liquid-gas case the
mass of the system is conserved whereas in this case the total number of
(LA+SA) states remains constant. The curve with $\gamma=.17$ has a clear
analogy with an isotherm above the critical point in the liquid-gas case.
However, there are shortcomings too. For instance, a constant $\gamma$ curve 
ends at the boundary on the LA side at a different frequency $\nu$ from the 
$\nu$ where the curve emerges out of the boundary on the SA side. Also, the 
boundary on the SA side has a curious unbounded upturn for all $\gamma<.12$. 
This peculiarity comes about because $\tau$ decreases along the SA side 
boundary as $\gamma$ is reduced from $\gamma=.12$ in Fig.3. An another
computational shortcoming in Fig. 5a is discussed in the last Section of this
paper. We are thus justifiably tempted to call the dynamical states to be in 
the LA and SA phases.
  
\begin{figure}[htp]
\centering
\includegraphics[width=15cm,height=10cm,angle=0]{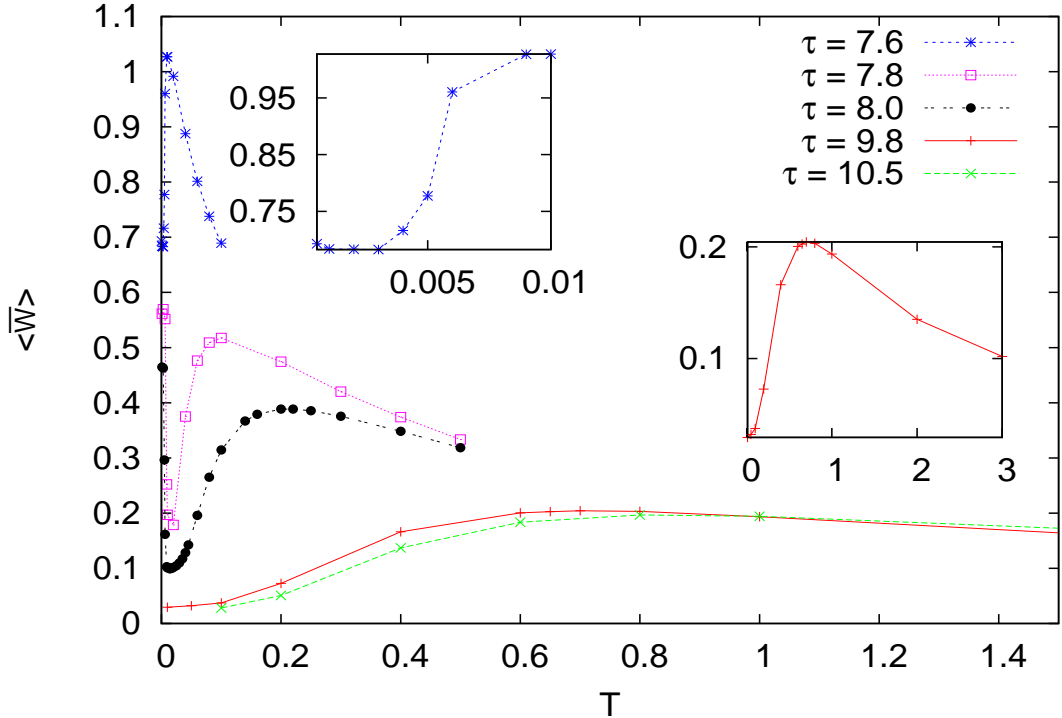}
\caption{The input energy $<\overline{W}>$ is plotted as a function of 
temperature $T$ for various values of the period $\tau$ of $F(t)$, for 
$F_{0}=0.2$ and $\gamma=0.12$. The curves are qualitatively separated into 3 
groups corresponding to small, intermediate and large $\tau$ values. The upper 
inset shows the variation of $<\overline{W}>$ for smaller $\tau$, indicating 
that there is no appreciable dip at low temperatures. The lower inset is a 
condensed version of large $\tau$ curves, showing the $<\overline{W}>$ peaks 
at large temperature $T\sim1.0$. The intermediate $\tau$ curves show typical 
SR behaviour with a characteristic dip at low $T$.}
\end{figure}

Fig. 6 shows the variation of $<\overline W>$ as a function of temperature 
for $\gamma=.12$ for a range of values of the period $\tau$ of the drive 
field. Though $<\overline W>$ shows peaking behaviour in all the 
graphs, these graphs can be classified into three qualitatively distinct 
groups. The three groups of $\tau$ can be identified by a careful look at the
insets of Fig. 6. For the group of graphs corresponding to the small $\tau$ 
values $<\overline W>$ peaks at very small $T$. At such small $T$ only 
intra-well transitions take place between SA and LA and no inter-well 
transitions are possible. Thus, as remarked in the introduction, this peaking 
of $<\overline W>$ cannot actually be termed as SR in a periodic potential. On 
the other hand, the graphs corresponding to the large $\tau$ values show a 
broad peak. They peak at temperatures $T>.5$. In this case, the peaking 
temperature is too large for the peak to be considered as a sensible SR. 
The graphs corresponding to the intermediate $\tau$ show the nature of typical 
SR with appreciably large inter-well transitions at the peak of 
$<\overline W>$ and a characteristic initial dip. This dip indicates that SA 
states are more stable than LA states for these parameter values. A detailed 
analysis shows that reasonable SR can be obtained in quite a narrow range of 
$\tau$ where the fraction of SA lies roughly between .3 and .7 at low 
temperatures. This range lies near about $\tau=8$ for almost all values of 
$\gamma$ between 0.07 and 0.16, Fig.3.

\subsection{The biharmonic drive}
 
From the above discussion it follows that physically reasonable SR can be 
obtained only for values of $\tau$ around 8, for this is the region where the 
two dynamical states occur in comparable fractions. We, therefore, consider a 
biharmonic drive with the main frequency close to $\omega=2\pi/8$. Henceforth, 
we take the period of the main component of the drive to be 8 unless otherwise 
stated. Therefore, in the case of biharmonic drive the harmonic component has 
the period of 4 and that of the subharmonic component a period of 16. Both 
these periods are far away from the range of periods where SA and LA occur in 
comparable fractions. Therefore, neither the harmonic component nor the 
subharmonic component alone is expected to contribute directly to the 
occurrence of SR. But their presence in the biharmonic (harmonic and 
subharmonic) drive case has considerable influence on the occurrence of SR and 
RE.

The presence of the harmonic component enhances the crest(trough) and distorts 
the (trough) (crest) of the main component. Nevertheless, the main component
($\omega \approx \omega_0$), being the fundamental component, is always 
reinforced by the harmonic component. Therefore, one would expect SR to occur
almost throughout the range of $f_m$ except at close to $f_m=0$. The 
subharmonic component, in this case being the fundamental component, 
successively enhances and reduces the alternate peaks (troughs) of the main 
component of $F(t)$ depending on the phase $\theta$. However, appreciable peaks
at the period of the main component appears only for large $f_m>.6$. Therefore,
one does not expect SR to occur at smaller values of $f_m$. However, one can 
expect RE, even if feeble, to occur throughout the range of $f_m$ except at 
$f_m=0,1$ for both harmonic and subharmonic drive cases.

Intuitively thinking, the ratchet current $\overline{v}$ should decrease if 
dissipation of energy of the system $<\overline{W}>$ increases given a drive 
field. This is what usually happens at large temperatures where both energy 
dissipation and ratchet current vary monotonically and $\overline{v}$ does not 
show current reversal. However, we are interested in studying the occurrence of
both SR and RE simultaneously. We may, therefore, also witness increasing
ratchet current with increasing dissipation.

\subsubsection{Harmonic field drive}

\begin{figure}[htp]
\centering
\includegraphics[width=15cm,height=10cm,angle=0]{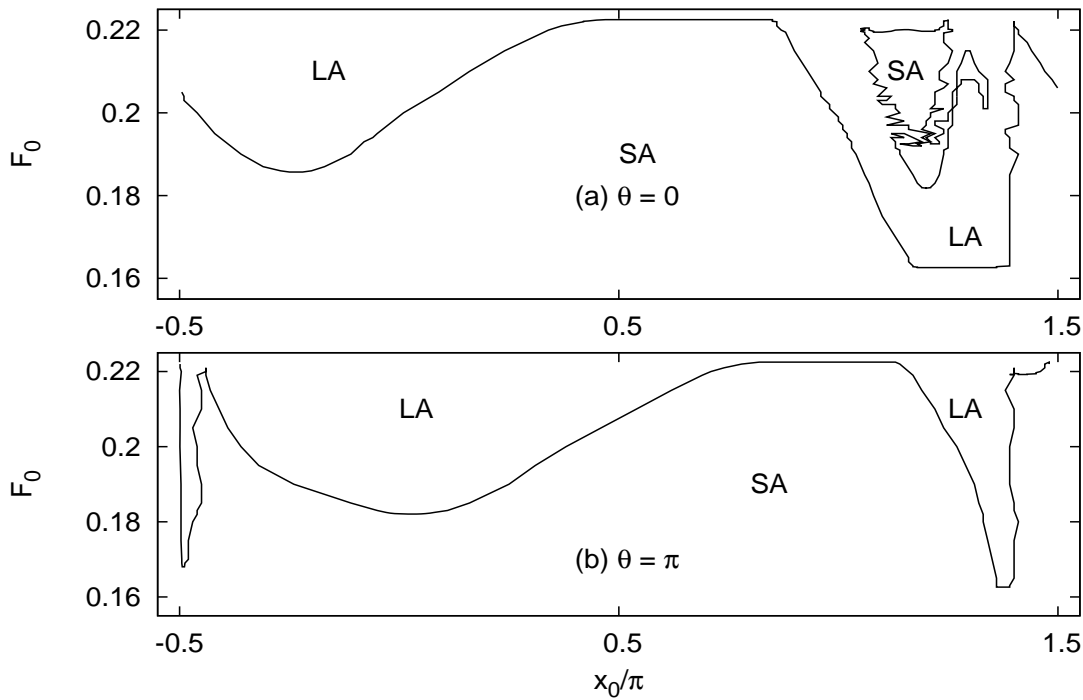}
\caption{shows the regions of LA and SA states obtained for initial
positions ${x_0}/\pi$ lying between -0.5 and 1.5, ie. within a full one
period of the potential for the initial phase $\theta$ of $F(t)$; $\theta=0$ 
(a) and $\theta=\pi$ (b), for $f_{m}=0.2$ and $\gamma=0.12$ at $T=0.000001$, 
and $\tau=8$.}
\end{figure}

\begin{figure}[htp]
\centering
\includegraphics[width=15cm,height=10cm,angle=0]{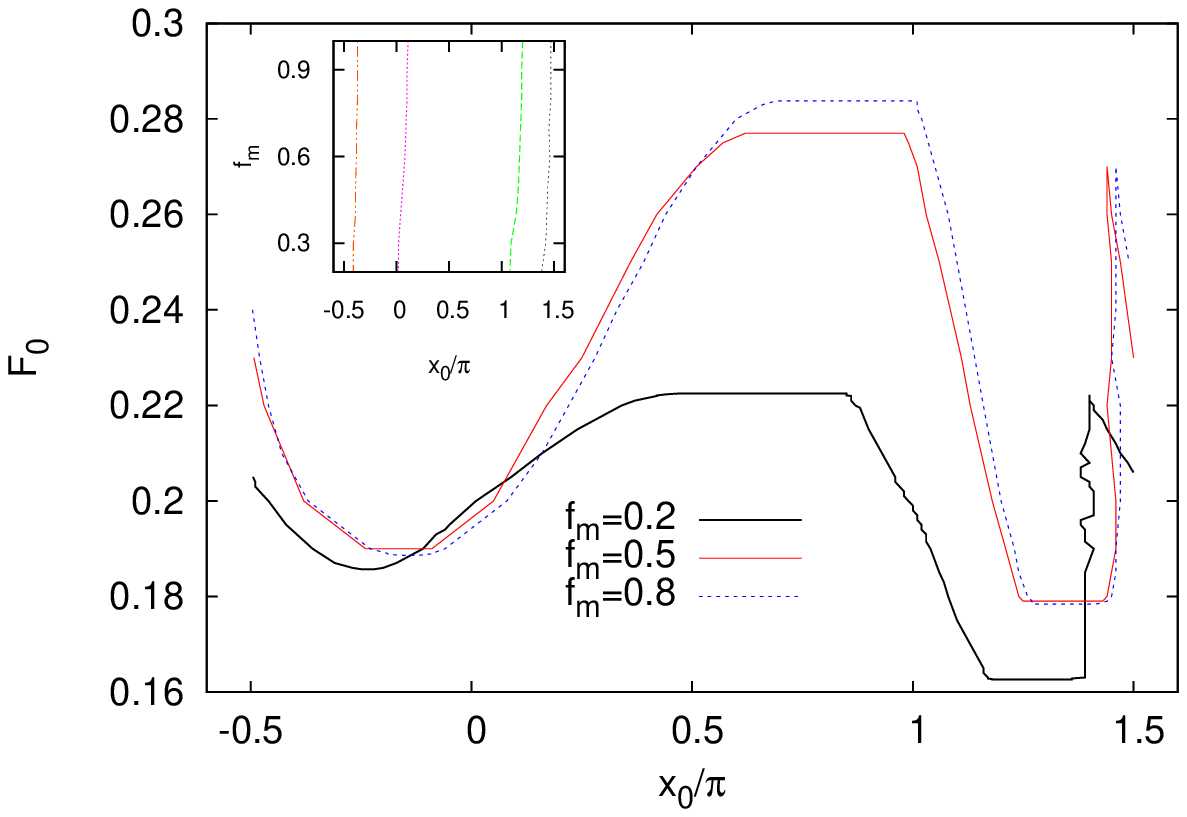}
\caption{shows the region of the states LA and SA, as in Fig.7, for
three values of $f_{m}=0.2,0.5$ and $0.8$, $\tau=8.0$, $\gamma=0.12$ at
$T=0.000001$. The inset shows how the boundary changes as $f_{m}$ is
varied for a fixed $F_{0}=0.2$.}
\end{figure}

Consider the external biharmonic field drive of the form given by Eq. (2.11).
In the following we take a fixed value of $\gamma=0.12$ and the period of
the main frequency term $\tau=2\pi/\omega=8$. In this case the harmonic 
component enhances the amplitude of the main frequency component. Fig. 1 gives 
the values of $\overline W$ corresponding to the states LA ($.17<F_0<.226$), 
SA ($F_0<.222$), LAH3 ($.217<F_0<.222$), LAH2 ($.226<F_0<.25$) and also when 
the system shows chaotic behaviour at large $F_0>.25$ values. Fig. 7a gives an 
idea of fractions of the state SA for various values of $F_0$ for 
$f_m=.2,~T=.000001$, and $\theta=0$. Fig 7b provides the same information 
as Fig. 7a but for $\theta=\pi$. In these figures no distinction has been made 
between LA and LAH3. Fig. 8 also gives the same information as Figs. 7 but
also shows how the SA boundary shifts as $f_m$ is changed for $\theta=0$. 
The informations provided in the Figs. 1, 7, and 8 are helpful in appreciating
the nature of variation of input energy $<\overline W>$. Since, SR does not 
occur for large $F_0$ values we restrict our studies to $F_0\leq.21$. 

\begin{figure}[htp]
\centering
\includegraphics[width=12cm,height=15cm,angle=0]{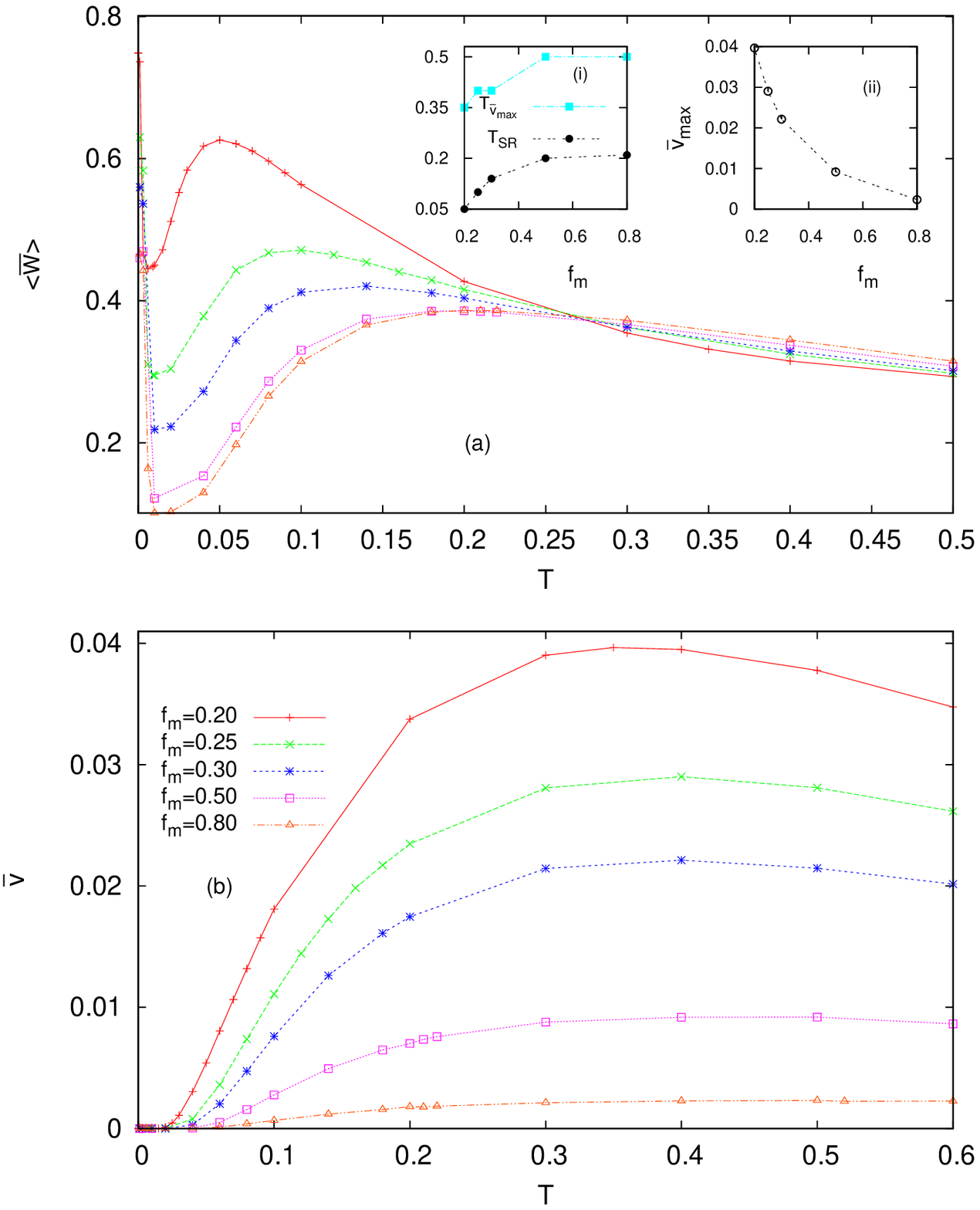}
\caption{The curves show the occurrence of SR (a) for $f_{m}=0.2, 0.25, 0.3,
0.5$ and $0.8$ for $\tau=8.0$, $F_{0}=0.2$ and $\gamma=0.12$. Fig. 9(b)
shows the variation of the ratchet current $\overline{v}$ again as a
function of temperature $T$ for the same parameter values. Inset (i) of (a)
shows the variation of $T_{SR}$, the temperature at which $<\overline{W}>$
 peaks, and
variation of $T_{\overline{v}_{max}}$, where $\overline{v}$ peaks as
a function of $f_{m}$ (obtained from (b)). The inset (ii) shows how
the peak value of $<\overline{W}>$ vary as $f_{m}$ is varied for a
 fixed value of $F_{0}=0.2$.}
\end{figure}

In Fig. 9, $<\overline{W}>$ (a) and the mean velocity $\overline{v}$ or the 
ratchet current (b), are plotted as a function of temperature for different 
values of $f_m$ and $F_0=0.2$ and $\theta=0$. The occurrence of both SR (Fig. 
9a) and RE (Fig. 9b) is evident. However, SR and the maximum of ratchet current 
$\overline{v}$ occur at widely separated temperatures (inset of Fig. 9a). 
Thus, the simultaneous occurrence of SR and RE is beyond dispute, though their 
connection is not clear. It is to be noted that the effective amplitude $F'_0$ 
for $F_0=0.2$ and $f_m=.2$ of the harmonic drive, Eq.(2.4), is 1.0. The 
amplitude is just at the threshold value making the potential barrier 
disappear momentarily every period in one direction whereas the barrier 
remains finite in the other direction throughout the period. For $F_0=0.2$ and 
$f_m<.2$ the potential barrier disappears for a finite interval of $t$ in a 
period whereas for $F_0=0.2$ and $f_m>.2$ the barrier always remains finite 
but unequal in the two directions. In either case this favours an asymmetric 
net particle motion with the assistance of fluctuating forces, but with no net 
impulse applied, resulting in RE. The presence of the harmonic component of 
$F(t)$ induces RE but, as explained earlier, it has only a supporting role to 
play for the occurrence of SR. 

The inset (ii) of Fig. 9b shows that the ratchet current rises fast as the 
fraction of the main component $f_m$ of $F(t)$ is decreased. This is because 
we have kept the parameter $F_0$ fixed at .2 for various $f_m$ values. However, 
decreasing $f_m$ corresponds to increasing $\alpha=\frac{f_h}{f_m}$ which also 
increases the effective amplitude $F_0'$ (Eqn. (2.7)) and hence $\overline{v}$.
The variation is particularly steep for $f_m<.2$ because $F_0'>1$ and hence 
the potential barrier disappears for a finite fraction of the period of $F(t)$ 
in one direction whereas the barrier remains large in the other direction.
Therefore, a different effect of $f_m$ will be revealed if $F_0'$ is kept 
fixed and $f_m$ is varied.  
 
\begin{figure}[htp]
\centering
\includegraphics[width=15cm,height=10cm,angle=0]{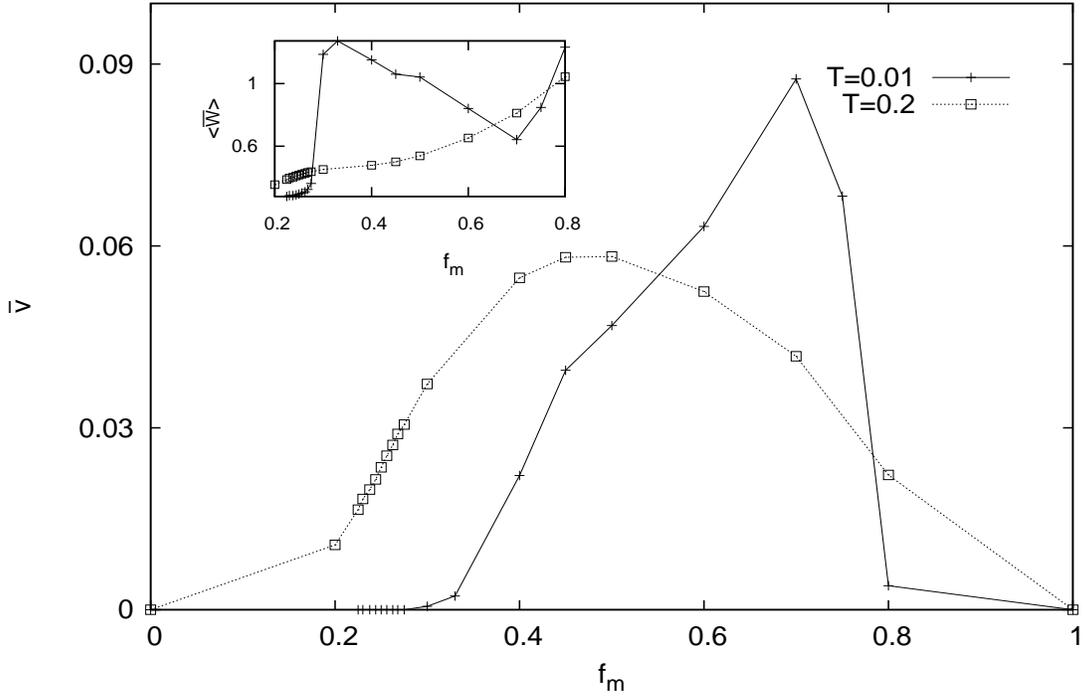}
\caption{The variation of $\overline{v}$ is shown as a function of $f_{m}$
 for a fixed value of $F_0'=\frac{F_{0}}{f_{m}}=0.8$ at temperatues
$T=0.01$ and $0.2$. The inset shows the corresponding variation of $<\overline{W}>
$.}
\end{figure}

Fig. 10 shows the variation of $\overline{v}$ as a function of $f_m$ for a
fixed value of $F_0'=.8$ for $T=.01$ and .2. The variation is as it should be 
because $f_m=0$ and 1 correspond to pure sinusoidal drive and hence there is 
no asymmetry in the system for a preferred direction, either to the right or 
to the left, of flow; $\overline{v}$ must be zero at these two extreme values 
of $f_m$ and nonzero at the intermediate values. However, for $F_0'=.8$, SR 
can be obtained only for a narrow range of $f_m$ that lies between $f_m=$ .2 
and .3 where ($.17<F_0\leq.21$). For comparison, we have plotted 
$<\overline{W}>$, in the inset, for the same parameter values. Only at large
$f_m$ values $<\overline{W}>$ and $\overline{v}$ vary such that one increases 
while the other decreases. But at or near the regions of occurrence of SR
dissipation $<\overline{W}>$ and the current $\overline{v}$ vary in the same 
way, at least for a part of the range of $f_m$ values.

\begin{figure}[htp]
\centering
\includegraphics[width=12cm,height=15cm,angle=0]{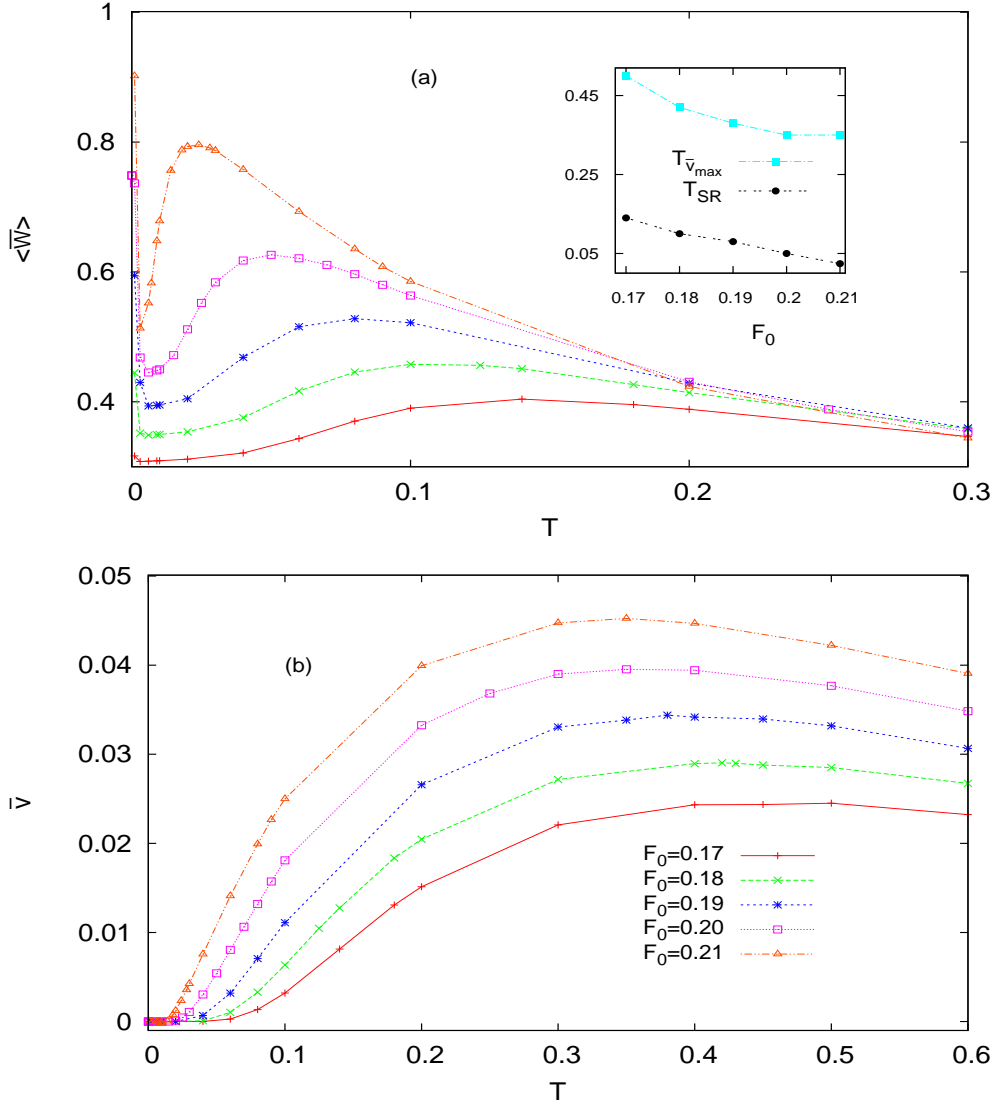}
\caption{Same as in Fig.9 (a) and (b) but for various values of $F_{0}$ ($=
0.17,0.18,0.19,0.20$ and $0.21)$ where SR occurs, for $f_{m}=0.2, \gamma
=0.12$ and $\tau=8.0$. Again the inset of (a) shows the variation
of $T_{SR}$ and $T_{\overline{v}_{max}}$ with $F_{0}$.}
\end{figure}

In Fig. 11, $f_m$ is taken equal to 0.2 ($\alpha=4$) and $<\overline{W}>$ (a), 
and $\overline{v}$ (b) are plotted for various $F_0$ values (.17, .18, .19, .2 
and .21). Note that for $F_0\geq .22$ no SR can be obtained. From these 
figures it is clear that in this range of parameters both SR and RE occur.
The inset of Fig. 11a again shows that $<\overline{W}>$ and $\overline{v}$
peak at widely separated temperatures, that is, they do not vary in total 
unison with the variation of temperature. Note that as $f_m$ is increased the
effective amplitude $F_0'$ decreases, correspondingly the inter-well 
transitions become less probable and hence $T_{SR}$ shifts to higher values. 

\begin{figure}[htp]
\centering
\includegraphics[width=12cm,height=12cm,angle=0]{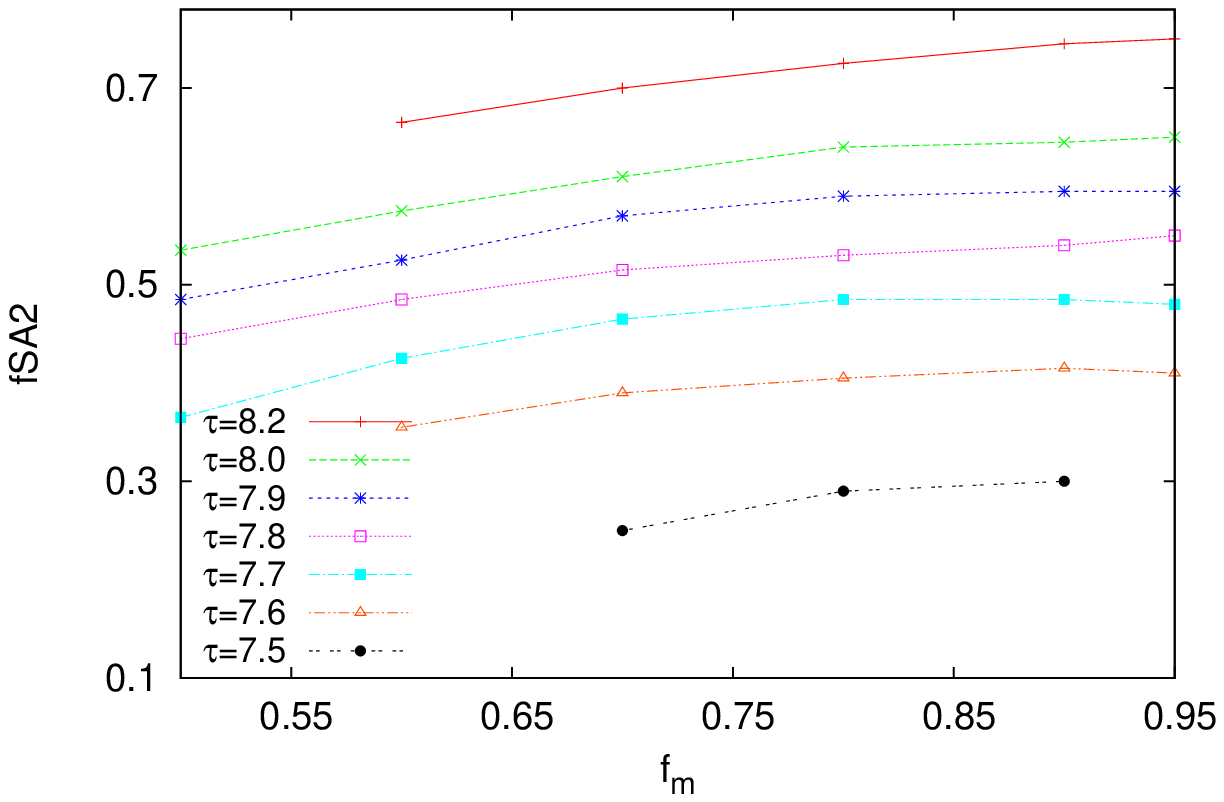}
\caption{The figure shows the variation of the fraction, fSA2 , of SA2 state,
as $f_{m}$ is changed for various values of $\tau$ and $F_{0}=0.2, \gamma
=0.12$ in the subharmonic drive case.}
\end{figure}

\subsubsection{Subharmonic drive}

\begin{figure}[htp]
\centering
\includegraphics[width=12cm,height=15cm,angle=0]{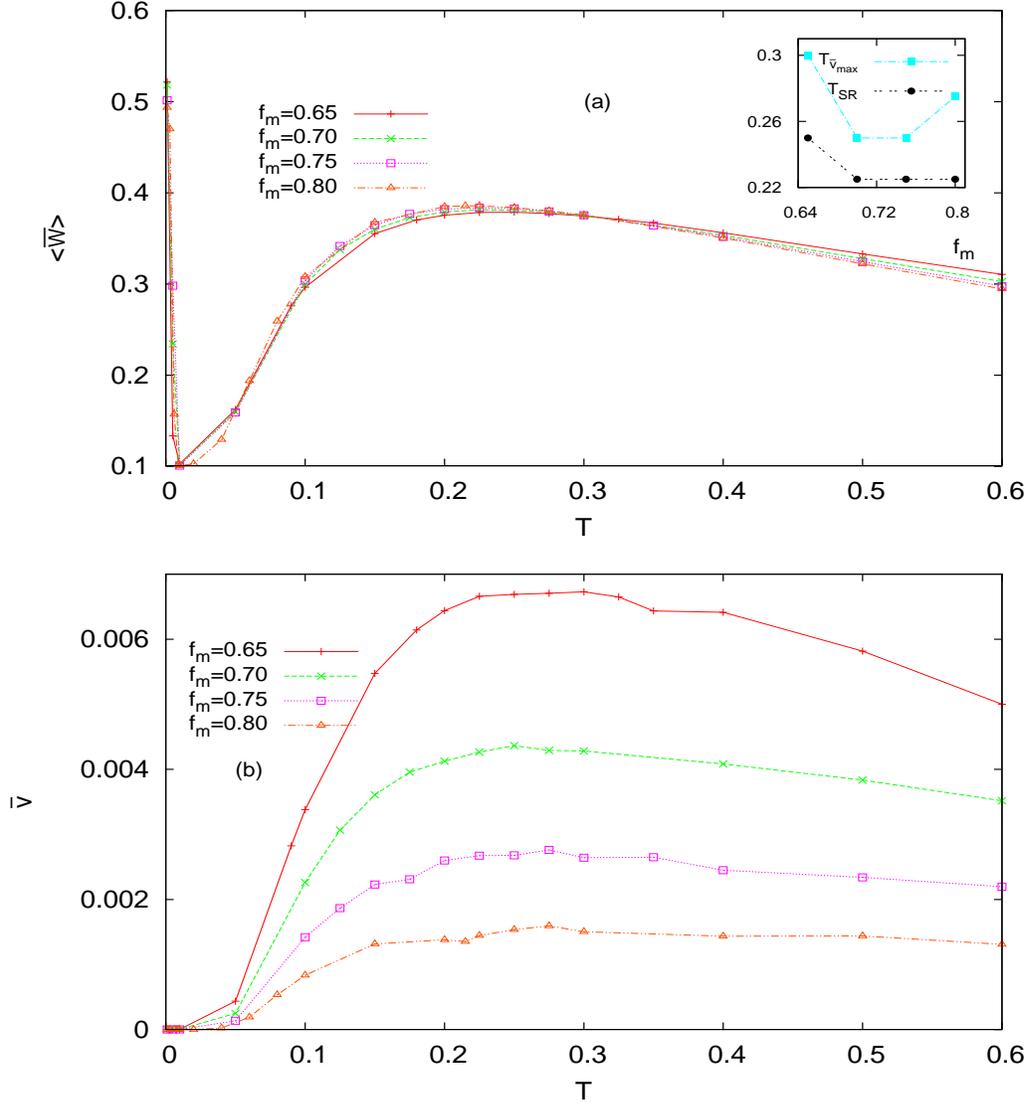}
\caption{Just as in Fig.9, $<\overline{W}>$ and $\overline{v}$ are plotted as
a function of $T$ for various values of $f_{m}$ in the subharmonic drive
case. The inset of the figure shows the locations $(T)$ of the maxima of
$W$ and $\overline{v}$ for $F_{0}=0.2,\gamma=0.12$ and $\tau=8.0$.}
\end{figure}

\begin{figure}[htp]
\centering
\includegraphics[width=12cm,height=15cm,angle=0]{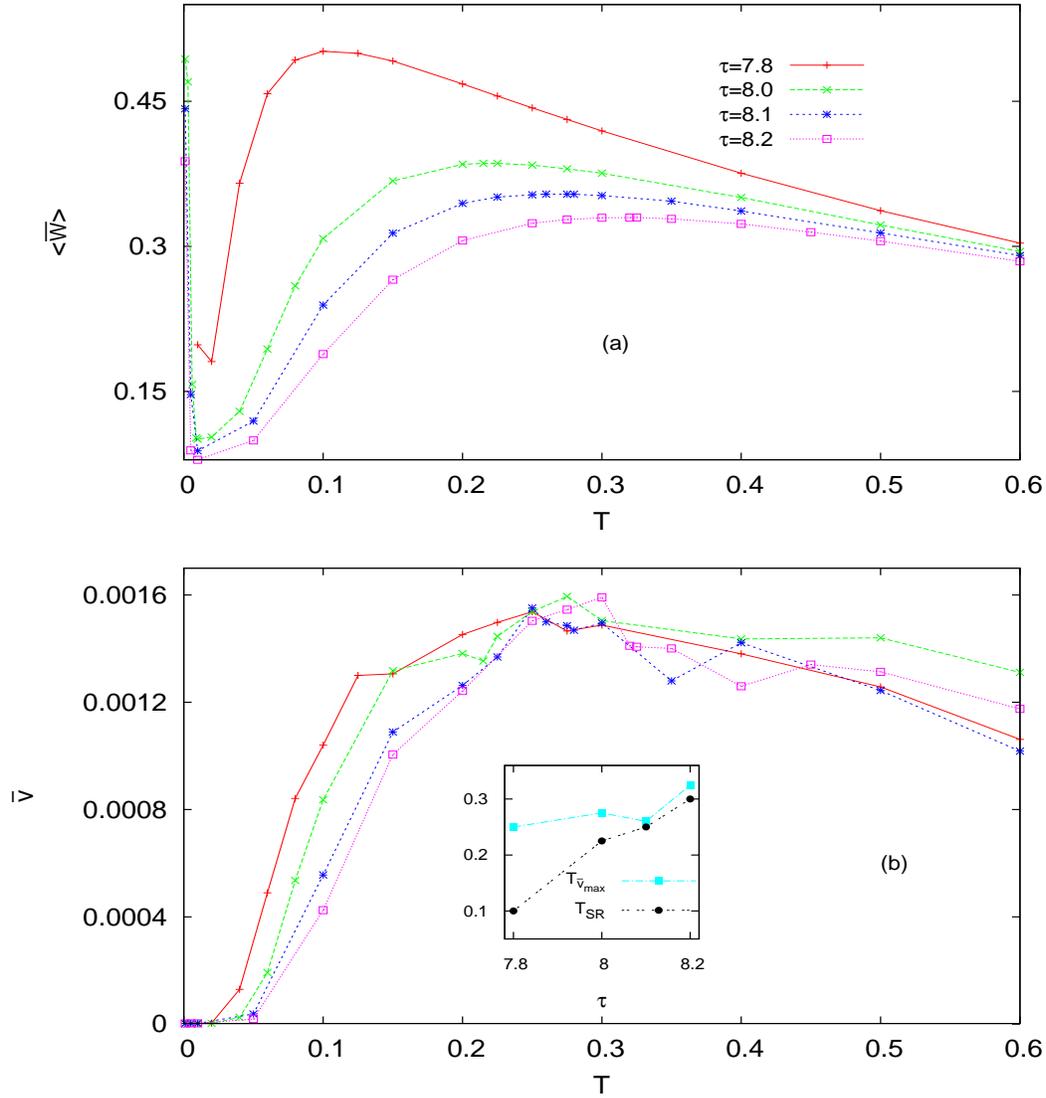}
\caption{Same as in Fig. 13 but for various values of $\tau$ and $f_{m}
=0.2,F_{0}=0.2$ and $\gamma=0.12$. The inset of (b) shows the variation of the
location of (along $T$ axis) of the maxima of $<\overline{W}>$ and 
$\overline{v}$.}
\end{figure}

\begin{figure}[htp]
\centering
\includegraphics[width=12cm,height=15cm,angle=0]{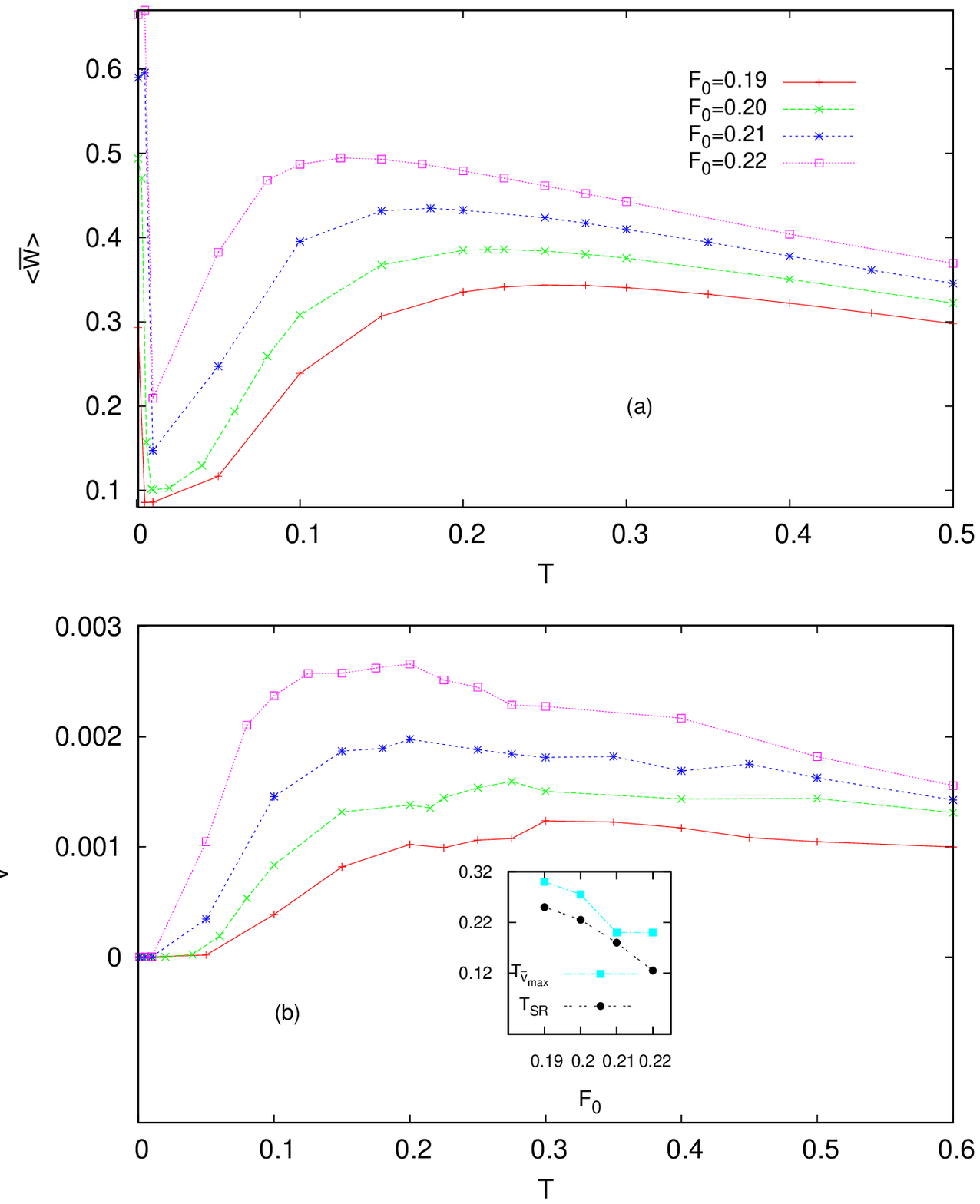}
\caption {$ <\overline{W}>$ and $\overline{v}$ are plotted as a function of 
temperature $T$ for various values of $F_{0}$ and $f_{m}=0.2,\tau=8.0$ and 
$\gamma=0.12$. The inset of (b) shows the location of the maxima of
$<\overline{W}>$ and $\overline{v}$ as $F_{0}$ is varied in the subharmonic 
case.}
\end{figure}

In this section we consider the subharmonic field drive, Eqn. (2.12). Naturally, 
it has a period of $\frac{4\pi}{\omega}$ representing the fundamental frequency
of $F(t)$ instead of $\frac{2\pi}{\omega}$ as in case of harmonic drive. Yet,
we call the term with period $\tau=\frac{2\pi}{\omega}$ as the main frequency 
term because we take $\omega\approx\omega_0$. As a consequence the particle 
trajectories obtained have a periodicity of $2\tau$ corresponding to the 
fundamental frequency of $F(t)$. These trajectories, again, are of two kinds 
representing the two dynamical states SA2 and LA2 in place of SA and LA, 
respectively,  of the harmonic drive case. Though the period of trajectories
corresponging to the states SA2 and LA2 is $2\tau$, the main frequency term 
with period $\tau$ is responsible for the occurrence of SR. The subharmonic 
component with frequency $\frac{\omega}{2}$ of the drive, however, helps in 
obtaining RE.

We restrict our calculations, in this section, to $\gamma=0.12$. Figure 12
provides a rough guideline for the proper choice of $\tau$ for various values 
of $f_m$ for $F_0=.2$. So far a sort of thumb rule has emerged that the best
region in the parameter space where SR can usually be obtained is where the
two dynamical states appear roughly in the same ratio (with a tolerance of 
less than 20 per cent) at low temperatures. The curves at the lower end of 
$f_m$ is abruptly ended because one of the dynamical states, namely LA2, 
disappears for all $f_m$ less than the edge of $f_m$ in the graphs.

We explore the occurrence of SR and RE in the region of parameter space where 
the occurrence of SA2 is around 50 per cent. In Fig. 13, we plot 
$<\overline{W}>$ (a) and $\overline{v}$ (b) as a function of temperature $T$ 
for $F_0=.2$, $\tau=8$ and $f_m=.65,~.7$, and .8. For $f_m<.65$ no reasonable 
SR could be obtained though $\overline{v}$ becomes large. For $f_m\geq .9$,
$\overline{v}$ becomes too small. Interestingly, the inset of Fig. 13a 
summarizes the result showing that $<\overline{W}>$ and $\overline{v}$ both 
peak at temperatures lying in a range between .2 and .25, though not exactly 
at the same temperature.

In Figs. 14a and 14b, respectively, show the variation of $<\overline{W}>$ and 
$\overline{v}$ with temperature for $F_0=.2$, $f_m=.8$ and $\tau=7.8,~8.0,~8.1$
and 8.2. One can observe that $<\overline{W}>$ peak at temperatures ranging
between .1 and .4 whereas $\overline{v}$ peak in a narrow range lying between
.25 and .3. Therefore for a specific $\tau$ both will peak at the same T value.
However, this will only be a happy coincidence and not a general rule that
both show peaking behaviour at the same temperatue or in a close range.

We again plot $<\overline{W}>$ and $\overline{v}$ as a function of temperature,
respectively, in Figs. 15a and 15b for $f_m=.8$, $\tau=8$ but now $F_0=.19,
~.2,~.21$ and .22. The peaks of $<\overline{W}>$ lie between $T=.25$ and .1. 
The peak for $F_0=.22$ is larger and sharper occurring at $T\approx .1$ We 
obtain a well defined SR for $F_0=.22$. It is to be noted that for $F_0=.22$ 
there was no SR in case of harmonic drive. The upper $F_0$ limit for the 
occurrence of SR is thus extended to a larger value in the subharmonic drive 
case. The $\overline{v}$ peaks are very broad for smaller $F_0$ values. It is 
very ambiguous to pinpoint the temperature at which $\overline{v}$ peaks, 
though there is a clear indication that the peaks lie between $T=.4$ and .1. 
That is, the peak temperatures of $<\overline{W}>$ and $\overline{v}$ tend to 
converge for larger $F_0$ values towards 0.1. We do not plot the graphs for 
$F_0<.19$ because $\overline{v}$ show relatively broader (almost flat) peak 
with much smaller heights. 

The results of the calculations show that SR and RE do occur in the same 
region of parameter space but a strong connection between the two phenomena is 
not conclusive.

\section{Discussion and Conclusion}

\begin{figure}[htp]
\centering
\includegraphics[width=15cm,height=10cm,angle=0]{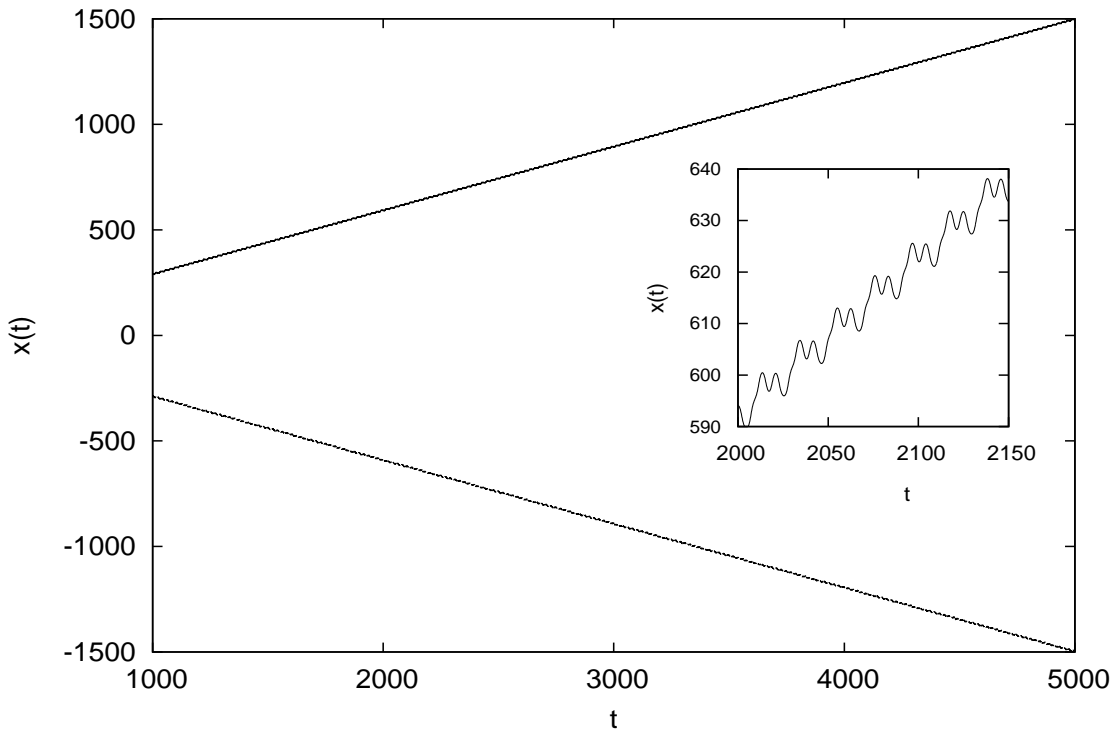}
\caption{The figure shows the particle moving forward (for initial condition
$x(0)=-.37\pi$) and backward (for $x(0)=-.36\pi$) with the same magnitude of 
constant average slope for $\gamma=0.04$ and $\tau=10.4$. The trajectory is 
actually periodically jerky as seen from the inset.}
\end{figure}

The diagram in Fig. 3 closes at around $\gamma=.17$ indicating that LA and SA
do not have separate identity for $\gamma\ge .17$. However, the diagram is
left open on the lower side of $\gamma$ at $\gamma=.07$. This is partly 
because of practical difficulty in clearly demarcating the regions of existence
of the two states. And partly also because the system's behaviour becomes
very complex at low values of $\gamma$. For example, for $\gamma=.04,~ \tau=
10.4$ and some initial $x(0)$ the trajectory of the particle becomes 
constantly (but periodically jerky) forward moving with a constant mean slope, 
Fig. 16. But for some other $x(0)$ the trajectory becomes backward moving with 
the same magnitude of slope. Note that the system is perfectly (right-left) 
symmetric and therefore these net transport giving trajectories can only be 
described as strange. However, when the total displacement is averaged over 
all the initial $x(0)$ values we get zero mean displacement as it should be 
for a pure sinusoidal drive.

The states LAH3 and LAH2 shown in Fig. 1 for the biharmonic drive at close to 
$T=0$ are, in fact, unstable against thermal fluctuations. As the temperature 
is raised from $T=0.000001$, by a small amount transitions takes place from 
these states to the LA state. This small raise in temperature is much smaller 
than the typical temperatures where $<\overline W>$ peak. Therefore, one can 
expect $<\overline W>$ to peak till $F_0\approx .22$ due to the transition
of SA to LA (not LAH3). For the small range
$.222<F_0<.226$ only LAH3 and LA appear at the lowest temperature. As the
temperature is raised LAH3 states quickly go over to LA states giving a small 
peak of $<\overline W>$. However, this is at a very low temperature where only 
intra-well thermal transitions are possible. Therefore, no genuine SR occurs in
this range of parameter space. For $F>.225$ the only state LAH2 existing at 
the lowest temperature also promptly goes over to LA as the temperature is 
raised by a little. Fig. 1, in fact, indicates that LAH2, as far as the 
$\overline W$ is concerned, is just a continuation of LA. From this maximum 
possible value $<\overline W>$ can only have a monotonic decrease because the 
trajectories become chaotic as the temperature is raised further. This can 
also be inferred from the observation that for a part of the period, $F(t)>1$ 
because the amplitude $F'_0>1$ for all $F_0>.2$ and $f_m=.2$.

The sharp decrease of energy dissipation $\overline W$ for the chaotic 
trajectories at large $F_0$ values appears surprising. However, the decrease
can be understood if the trajectories are studied carefully. In the usual 
situations energy is absorbed from the external field if the particle moves
against the force. However, energy absorption becomes negative if particle
moves in the direction of the force. In the case of chaotic trajectories the
latter situation becomes quite common and hence the decrease of energy 
absorbed from the field at large values of $F_0$ and also at large 
temperatures.

In Fig. 5a, the constant $\gamma$ curves appear to be intruding into the 
coexistence region on the SA side. That is only because of our compuational 
inadequacies in pinpointing the coexistence boundary. The LA states disappear 
gradually and very slowly. Moreover, the appearance of states (LA or SA) 
depend very sensitively on the initial condition ($x(t=0)$). We take at most
200 discrete values of the positions within a period of the potential. This 
cannot be considered as a continuous scanning. There is every possibility of
missing an LA state and set the boundary before we have actually reached 
there (fSA=1.0, for the first time). The $\gamma=.16$ curve should really be 
grazing close along the actual boundary. The correspondence with the liquid-gas 
phase diagram is reasonable.

Finally, to summarise in conclusion, the occurrence of SR in a periodic 
potential when driven by a biharmonic field of appropriate frequency is shown
to be possible. The parameter range in which SR occurs is quite narrow but 
finite. RE too occurs in this range as it does outside of it. Thus, SR and RE
do occur in the same parameter space. However, the peak of $<\overline W>$ and 
the maximum ratchet current  occur at temperatures widely separated from  
each other in the harmonic drive case. In the subharmonic drive case, however,
there can be found exceptional circumstances where the maxima of the two 
quantites may occur at the same temperature. But these are mere coincidences. 
The occurrence of the two phenomena at the same parameter range can allow for 
efficient material transport at optimum tempratures at certain frequencies of 
field drive in periodic potential systems. However, at that optimum tempearture particle current need not be a maximum.    

Partial financial support from UGC, India under the Special Assistance Program 
is acknowledged.

\end{document}